\documentstyle[11pt,aaspp4]{article}
\def\puncspace{\ifmmode\,\else{\ifcat.\C{\if.\C\else\if,\C\else\if?\C\else%
\if:\C\else\if;\C\else\if-\C\else\if)\C\else\if/\C\else\if]\C\else\if'\C%
\else\space\fi\fi\fi\fi\fi\fi\fi\fi\fi\fi}%
\else\if\empty\C\else\if\space\C\else\space\fi\fi\fi}\fi}
\def\SP{\let\\=\empty\futurelet\C\puncspace }

\def\etal{et\SP al.\SP }

\def\deg{$^\circ$\ }
\def\kms{kms$^{-1}$}
\def\eg{$e.g.$,\SP}
\def\h1{$h^{-1}$}
\newcommand{\simless}{\stackrel{\scriptstyle <}{\scriptstyle \sim}}

\def\lsim{~\rlap{$<$}{\lower 1.0ex\hbox{$\sim$}}}
\def\gsim{~\rlap{$>$}{\lower 1.0ex\hbox{$\sim$}}}
\def\void#1{{}}

\begin{document}

\title{The Southern Sky Redshift Survey \footnote{Based on observations at 
Cerro Tololo Interamerican Observatory (CTIO), National Optical
Astronomy Observatories (NOAO) which is operated by the Association of
Universities for Research in Astronomy, Inc. under contract to the
National Science Foundation; Complejo Astronomico El Leoncito (CASLEO),
operated under agreement between the Consejo Nacional de
Investigaciones Cient\'\i ficas de la Rep\'ublica Argentina and the
National Universities of La Plata, C\'ordoba and San Juan;
European Southern Observatory (ESO), partially under the bilateral agreement
ESO-ON; Fred Lawrence Whipple Observatory; Laborat\'orio Nacional de
Astrof\'\i sica; and the South African Astronomical Observatory }}

\author{L. Nicolaci da Costa,\altaffilmark{2}}
\affil{European Southern Observatory, Karl Schwarzschild Str. 2, 
85748 Garching-bei-M\"unchen, Germany}

\author {C. N. A. Willmer,
P.S. Pellegrini, O. L. Chaves,
 C. Rit\'e, M. A. G. Maia } 
\affil{Departamento de Astronomia, Observat\'orio Nacional, Rua
General Jos\'e Cristino 77, Rio de Janeiro, 20921-030, Brazil}

\author{M.J. Geller, D.W. Latham, M.J. Kurtz, J.P. Huchra}
\affil{Harvard-Smithsonian Center for Astrophysics, 60 Garden St.,
Cambridge MA 02138 }

\author{M. Ramella,}
\affil{Osservatorio Astronomico di Trieste, Via G. B. Tiepolo 11,
34131 Trieste, Italy}

\author{A.P. Fairall,}
\affil{Department of Astronomy, University of Cape Town, 7735
Rondebosch, South Africa}

\author{C. Smith,}
\affil{Department of Astronomy, University of Michigan, Ann Arbor, MI}

\and

\author{S. L\'\i pari}
\affil{Observatorio Astron\'omico de C\'ordoba, Laprida 854,
C\'ordoba, 5000, Argentina}

\altaffiltext{2}{Departamento de Astronomia, Observat\'orio Nacional,
Rua General Jos\'e Cristino 77, Rio de Janeiro, 20921-030, Brazil}

\begin{abstract} We report redshifts, magnitudes and morphological
classifications for 5369 galaxies with $m_B \leq 15.5$ and 57 galaxies
fainter  than this limit, in two regions covering a total of 1.70
steradians in the southern celestial hemisphere. The galaxy catalog is
drawn primarily from the list of non-stellar objects identified  in the
Guide Star Catalog (Lasker et al. 1990, AJ 99, 2019; hereafter GSC). 
The galaxies have positions accurate to $\sim 1^{\prime\prime}$ and
magnitudes with an rms scatter of $\sim 0.3^m$. We compute magnitudes
($m_{SSRS2}$) from the relation  between instrumental GSC magnitudes and
the photometry by Lauberts \& Valentijn (1989). From a comparison with
CCD photometry, we find that our system is homogeneous across the sky
and corresponds to magnitudes measured at the isophotal level $\sim$ 26
mag arcsec$^{-2}$. The precision of the radial velocities is of $\sim$
40 \kms\ and the redshift survey is more than  99\% complete  to the
$m_{SSRS2}$ = 15.5 magnitude limit.  This sample is in the direction
opposite to the CfA2; in combination the two surveys provide an
important database for studies of the properties of galaxies and their
large-scale distribution in the nearby Universe.

\end{abstract}
\clearpage

\section{Introduction}

During the past two decades, complete, wide-angle redshift surveys have
provided a fundamental basis for exploring the large-scale distribution
of galaxies. Starting with the  CfA1 Redshift Survey (Davis \etal 1982),
the first to reach beyond the Local Supercluster, these surveys have
revolutionized our view of the spatial distribution of galaxies. The
CfA1 redshift survey provided the first three-dimensional glimpse of the
nearby universe, showing that within 70$h^{-1}$ Mpc ($H_0$ =100\h1
km/s/Mpc) the galaxy distribution is far from homogeneous with large
regions devoid of bright objects. The first complete slice of the CfA2
Redshift Survey (de Lapparent et al. 1986) soon produced striking
results with the discovery of the Great Wall, a spectacular example of a
well-defined, thin, two-dimensional structure perpendicular to the
line-of-sight and  outlining several large voids (de Lapparent et al.
1986). Later the diameter-limited Southern Sky Redshift Survey (SSRS, da
Costa et al. 1988), not only confirmed the existence of large voids, but
also detected other thin two-dimensional coherent structures, like the
Southern wall.  These surveys revealed the large scale of structure in
the universe.

Surveys of both hemispheres have continued: the CfA2 covers a portion of
both galactic caps in the northern celestial hemisphere (Geller \&
Huchra 1989) and the SSRS2 (\eg da Costa et al. 1994$a$) covers a
portion of both caps in the south. In contrast to the 
diameter-limited SSRS, the deeper SSRS2 is a magnitude limited survey
based on a selection of objects from the Hubble Space Telescope Guide
Star Catalog (Lasker \etal 1990; GSC hereafter).

The CfA2 and SSRS2 surveys now cover more than 30\% of the sky. Together
they contain more than 20,000 galaxies and provide a panoramic view of
the three-dimensional galaxy distribution  (\eg da Costa \etal 1994$a$).
The redshift survey maps show that the galaxy distribution consists of a
complex network of voids, typically 5000 \kms in size, surrounded by
thin walls where most galaxies reside (Geller \& Huchra 1989).  A
quantitative analysis of  the SSRS2 (El-ad \etal 1996) confirms this
picture, finding the typical size of voids to be $\sim$ 40\h1 Mpc. 
Contrary to some earlier  expectations, large voids and walls are not
rare events; they are  present in every direction surveyed (\eg CfA2;
Las Campanas Redshift Survey, Schectman et al. 1996). 

Both separately and combined, the CfA2 and SSRS2 have provided a basis
for determination of the local luminosity function (Marzke \etal
1994$b$, da Costa \etal 1994$a$) and its dependence
on morphology (Marzke \etal 1994$a$, Marzke \etal 1998) and color
(Marzke \& da Costa 1997); the power-spectrum of the galaxy distribution
(Park \etal 1994, da Costa \etal 1994$b$); the mean pairwise velocity
dispersion (Marzke \etal 1995); the dependence of clustering properties
on luminosity (Benoist \etal 1996), morphology and color (Willmer et al.
1998) and the properties  of loose and compact groups (Barton \etal
1996, Ramella \etal 1997, Ramella \etal 1998).

Even though the currently planned deeper surveys will dramatically
extend the depth of the sampled volume of the Universe, the combined
CfA2 and SSRS2, which probe the distribution of $L_*$  galaxies out to
$\sim$ 100 \h1 Mpc, will continue to be a valuable database. For
instance, the large number of galaxies, the complete sampling and the
large angular coverage will allow, in conjunction with independent
distance measurements, a more detailed study of the relation between
individual galaxies and large-scale structure.

Here we provide the SSRS2 catalog. In Section 2 we discuss the assembly
of the sample of galaxies from the GSC, the magnitude system, and  the
morphological classification of galaxies. In Section 3 we describe the
observations. The catalog itself is in Section 4. A brief summary
follows in Section 5. 

%
%
\section{The Sample}

\subsection{Constructing the catalog}

We selected the SSRS2 galaxy sample from the list of non-stellar objects
in the GSC.  The sample covers the region --40\deg $\leq \delta \leq$
--2.5\deg and b $\leq$ --40\deg in the Southern Galactic cap (SSRS2
south), and the region $\delta \leq 0$\deg and b $\geq$ +35\deg  in the
Northern Galactic cap (SSRS2 north). The northern limits in declination
make the SSRS2 contiguous with the CfA2.

The GSC has good astrometric accuracy ($\sim$ 1$^{\prime\prime}$) and
objectively measured instrumental magnitudes.  To examine the adequacy
of the instrumental GSC magnitudes, we identified the GSC counterparts
for about 1000 galaxies from the Lauberts and Valentijn (1989, hereafter
ESO-LV) catalog, and compared the instrumental and ESO-LV magnitudes. We
find a linear relation with small scatter ($\approx$ 0.3$^m$) between
these magnitudes. This relation yields  a transformation between the GSC
instrumental magnitudes and the ESO-LV measurements (Alonso \etal 1993).
Using this relation (see Section 2.2) we extracted a preliminary sample
of objects from the GSC with m$_{ssrs2} \leq 15.5$.

The ``non-stellar'' GSC list includes many visual double or triple
stars, parts of bright galaxies, and satellite and asteroid trails.
Thus it is not possible to equate all ``non-stellar'' objects with
galaxies. To identify the galaxies in this preliminary sample, we
either matched the list with other known galaxy catalogs or visually
inspected the images. The typical number of ``non-stellar'' objects in
a $10^\circ$ declination strip of the GSC, within the survey boundaries,
is $\sim$ 5500; only 900 ($\sim$ 16\%) of these are galaxies.

To investigate the nature of the selected objects, we  matched the list
with the following series of galaxy catalogues: 1) The ESO-LV catalog
yielded 1477 galaxies in common with the SSRS2 south and  338 in the
SSRS2 north for declinations $\delta \leq -17.5^\circ$, the  northern
limit of the ESO-LV catalog. For a match, we required a separation
$\simless 3^{\prime}$; 2) For declinations  $\delta > -17.5^\circ$ we
matched the sample with the Morphological  Catalogue of Galaxies
(Vorontsov-Velyaminov \& Arkhipova 1963-1968,  hereafter MCG) allowing
separations $\leq$4$^{\prime}$ (because of the  large errors in the
catalog coordinates) to find 900 matches in the SSRS2 south and 837 in
the SSRS2 north; 3) We then matched the remaining objects  to  a subset 
of the APM survey with $m_B \simless 16$ (Loveday 1996) in the
southern galactic cap  ($-40^\circ \leq \delta \leq -2.5^\circ$, kindly
provided by S. Maddox  and W. Sutherland).  We cross-identified $\sim$
700 APM galaxies. For  the SSRS2 north, a similar match to an APM
catalog (Raychaudhuri \etal 1994, kindly provided by Somak Raychaudhuri) 
yielded $\sim$580 matches.

Taken together the matching procedures identified  $\sim$ 88\% of the
galaxies in the final catalog.  The matching procedure also showed that
galaxies with very large apparent diameters are not included in the GSC,
but can be found in either the ESO-LV or the MCG. In these cases, we
used the coordinates from the  parent catalog and converted the
magnitudes using the appropriate  relation (see Section 2.2).  We
recover 124 galaxies from the ESO-LV  and 173 from the MCG. 
These large objects are missing in the APM  catalog as well as in the
GSC ; both suffer from the same bias.  These cases constitute $\sim$ 6\%
of the sample.

Finally, we visually inspected the remaining unmatched  objects in the
preliminary  GSC "non-stellar" list. We identified $\sim$ 190 galaxies
in the southern and 100 in the northern subsamples accounting for
$\sim$6\% of the final sample. Visual inspection revealed that a few
($\sim$ 30) bright and single galaxies were split in the GSC into two or
more components. In these cases we maintained the object as a single
entry in the catalog and used magnitudes from the ESO-LV or MCG
catalogs, converted into the $m_{SSRS2}$ system (Section 2.2).  Visual
inspection also showed  that quite often close pairs of galaxies
(especially galaxies classified  as ``interacting galaxies'' (IG) by
Lauberts (1982)) appear as a single object in the GSC.  If we had
detailed information for each galaxy from the ESO-LV (coordinates,
magnitudes and velocities) we used it to split the GSC entry  into the
appropriate number of galaxies, with  appropriate coordinates and
magnitudes converted to the $m_{SSRS2}$  system.  In cases where data
for the separate galaxies were unavailable,  we estimated the
magnitudes.  For about 16 galaxies, the estimated magnitude was brighter
than our magnitude limit and new entries were created in the catalog, and
new coordinates measured.  We believe that this careful inspection
insures the completeness of the catalog to the desired limiting
magnitude.

As a final note, we point out that some galaxies may still be missed
depending on the performance of the galaxy/star classification algorithm
used by the GSC.  In order to evaluate this effect we searched for APM
galaxies in the GSC ``stellar'' list, finding that about half of the
matches had already been included in the SSRS2 catalog from our searches
in the ESO and MCG catalogs.  The results indicate that our 
incompleteness, due to the misclassification of galaxies as stars, is
$\lsim$ 2\%,  occuring primarily in the fainter 0.5 magnitude bin.

%
%
\subsection{Magnitudes}

The SSRS2 magnitude system converts the GSC instrumental magnitudes to 
the magnitude system of the ESO-LV catalog.  About 1000 GSC 
``non-stellar'' objects matched to ESO-LV galaxies determine the linear 
relation between the GSC instrumental magnitudes, $m_{GSC}$ and  ESO-LV
$B_T$ magnitudes.  We then add a constant term to these $B_T$ 
magnitudes to match the Zwicky-B(0) (Huchra 1976; de Vaucouleurs \&  de
Vaucouleurs 1964) system of the CfA2. From Felten (1985), we take a 
constant term of 0.26 $^m$. The relationship between $m_{SSRS2}$ and 
$m_{GSC}$ is then (Alonso \etal 1993) \begin{equation} m_{SSRS2} = 0.59\
m_{GSC} + 8.42 \end{equation}

A comparison of the $m_{SSRS2}$ magnitudes with CCD photometry (Alonso
\etal 1993, 1994) shows that the $m_{SSRS2}$'s  have a limiting 
isophotal level close to 26 mag arcsec$^{-2}$.  The rms uncertainty in 
the magnitudes is $\sim$ 0.30$^m$. Alonso \etal (1994) also verify that 
the photometry is uniform across the survey. They found a mean offset
{\bf $m_{B}(26)$ -- $m_{SSRS2}$=} --0.02;  the variation of the
zero-point over the sky is $\sim\pm$ 0.10 magnitudes for regions at
large angular  separations.  From a small number of galaxies for which
both $m_{SSRS2}$  and $m_{Zwicky}$ are available, Alonso \etal (1994)
also show that the zero-point offset is $m_{SSRS2}$ -- $m_{Zwicky}$
$\sim$ 0.10$^m$, smaller than the typical magnitude errors in either
catalog (0.35$^m$ for $m_{Zwicky}$). 

To further justify the above results in figure 1 we compare
$m_{SSRS2}$ magnitudes, as defined above, with $B_{26}$ taken from the
ESO-LV for 1588 galaxies in common. In the figure, we also show the
linear $\chi^2$ fit obtained in the domain $13 \le m_{SSRS2} \le
15.5$, which yields the relation $B_{26}$ = 1.01 $m_{SSRS2}$ - 0.10,
with a dispersion of about 0.33$^m$. Assuming the errors in $B_{26}$
to be $\sim$ 0.1$^m$, we estimate the errors in $m_{SSRS2}$ to be
about 0.31$^m$, confirming the findings of Alonso \etal (1994), for a
much larger sample of galaxies.

For single objects which were split in the GSC, or for two or more
galaxies with a single entry in the GSC, we use magnitudes tabulated
in the ESO-LV or MCG catalogs converted into $m_{SSRS2}$ by adding
0.26$^m$ to the ESO-LV $B_T$ or by adding 0.5$^m$ to the MCG
magnitude. For merged images not separated by the GSC, and not
contained in the ESO-LV nor in the MCG, we use the GSC magnitude to
obtain $m_{SSRS2}$, as given by equation (1), rather than including a
visual estimate.

Figure 2 shows the number counts of galaxies in 0.5 magnitude  intervals
for both subsamples (North and South) of the SSRS2. Although there are
small sample fluctuations for bright galaxies ($m_{SSRS2} \simless 11$),
the counts for the two subsamples are in excellent agreement at fainter
magnitudes and both have a slope of  $\sim$ 0.6.  There is no flattening
of the slope at the faintest magnitudes, indicating that there is no
significant loss of objects at the limit of the survey.

%
%
\subsection{Morphology}

Although the visual classification of a ``non-stellar'' object as a
galaxy is straightforward, the morphological classification is not (\eg
Lahav \etal 1995); classifications vary from individual to individual.
In  the SSRS2 catalog, we include morphological classifications for all
galaxies available in the ESO-LV for $\delta \leq -17.5^\circ$ (1815 
galaxies); from the ZCAT compilation of Huchra (1996, private 
communication); and from an APM classification (W. Sutherland, private 
communication) in the declination range  $-10^\circ \leq \delta \leq
-2.5^\circ$ for the southern galactic cap only.  

The ESO-LV morphology is the most detailed and homogeneous
classification, subdivided into transition types and subclasses, and
we adopt it as our reference system.  The APM classification includes
only principal types and we transform them into T types: ellipticals
(T=-5), S0s (T=-2), spirals (T=5), and irregulars (T=10).  ZCAT
morphologies are not so homogeneous as the other catalogs because it
is a compilation which includes data from several sources. We also map
these classifications to conform with the ESO-LV system. We point out,
however, that while in the ESO-LV system, T=5 corresponds to generic
spirals, here we have assigned this type for all spirals pending more
detailed classification. We also devised numerical codes for peculiar
galaxies (T=15) and unclassified objects (T=23).

To provide complete morphological information, we visually inspected
galaxies with no previous classification.  To calibrate our
classifications and to make them as consistent as possible with the
ESO-LV system we re-classified galaxies in the ESO-LV without looking
at the published T type.  For the SSRS2 south we used film copies of
the ESO(B) Atlas and, in doubtful cases, we also used the ESO-SRC (J)
film copies.  By iteratively classifying and checking, we became
confident in our classifications when they matched the ESO-LV more
than 85\% of the time.  Most of the discrepancies come from
misclassifying S0s, either as ellipticals or as early spirals. We then
used the ESO (B) film copies to classify objects south of declination
$\delta = -17.5^\circ$; northward of this limit we used paper copies
of the Palomar Observatory Sky Survey (POSS). For the SSRS2 north,
which was examined later, we used the Digitized Sky Survey.  The lower
photographic quality of the POSS paper copies may have led us to
misclassify some early spirals as SOs or even ellipticals.  However,
we classified only about 200 galaxies ($\sim$ 4\% of the sample) from
the POSS.

We emphasize that our system of morphological types is only homogeneous
and complete when coarsely binned, $i.e.$, if we define ellipticals as
T=[-5 to -3], lenticulars as T=[-2 to 0], and spirals as T=[1 to 9].
Adopting this definition we find that the magnitude-limited SSRS2
consists of: 12.5\% of ellipticals, 19.0\% of S0s,  64.0\% of spirals,
1.5\% of irregulars, 2.6\% of peculiars and 0.4\% of unclassified
galaxies. More recently we began using images from the  Digitized Sky
Survey and the process of classifying all objects with greater
morphological resolution is underway.  We will periodically update the
catalog version available at our www site mentioned below.

Figure 3 shows the fraction of early and late type galaxies in  0.5
magnitude bins. While at the bright end ($m_B \simless 13.5$) there are
fluctuations, due to small number statistics, at fainter magnitudes, the
population fractions approach a constant value of about  66\%  and 32\% 
for late and early type galaxies, respectively.

%
%
\section{Observations}

The acquisition of spectra for the SSRS2 began in 1987. We accumulated
data with a variety of telescopes and instrument setups until 1997. In
Table 1 we list the different sites, telescopes and detectors.
We also list the slit size, wavelength
coverage, and the mean resolution of spectra.

In the first observing runs at CASLEO and all observations at Mount
Hopkins and the SAAO, we used Reticon photon-counting devices. 
The CTIO, ESO, LNA and the 1996 CASLEO spectra are all CCD data.

We reduced the Reticon spectra following the standard procedure
(Tonry \& Davis 1979). We combined the spectra of each object and then
subtracted the sky counts which we observed simultaneously in a separate 
slit.  We then flatfielded the sky-subtracted spectrum to remove detector
pixel-to-pixel response. We wavelength calibrated the flatfielded spectrum.
The typical error in the polynomial fit for wavelength calibration is 
$\sim$ 0.3 to 0.4 \AA. 

We also reduced the CCD spectra in a standard fashion (\eg Massey 1992)
using IRAF routines. We bias-subtracted, flatfielded, and when
necessary, corrected the raw frames for illumination effects. We then
extracted the object and comparison spectra using the optimal extraction
routines available in IRAF. We did the wavelength using the REDUCER
package (provided by W. Wyatt of the CfA), a C language adaptation of
the FORTH routines used to reduce the Reticon spectra.

We used the standard cross-correlation technique (Tonry \& Davis 1979)
and emission line fitting to extract redshifts.  For the first CASLEO
runs, we used the FORTH version of the cross-correlation
code. However, we measured most of our radial velocities using the
RVSAO package (Kurtz \etal 1992) in an IRAF environment.  We used only
two of the default templates supplied with RVSAO: a composite stellar
spectrum and a composite galaxy spectrum. Whenever we detected
emission-lines, we also used them to measure the redshift. If both
cross-correlation and emission-lines provided a radial velocity, we
combined these measurements with weights proportional to the estimated
uncertainty in each determination (\eg Tonry \& Davis 1979).  For
velocities with estimated errors below 20 \kms, usually based on
emission lines, we added 25 \kms in quadrature to account for other
sources of systematic errors as suggested by Kurtz and Mink (1998).
The internal error in most of our velocities is $\sim$ 40 \kms (e.g. da
Costa \etal 1984, da Costa \etal 1991).

%
%

\section{The Catalog}

The number of galaxies in the survey is 5426: 3489 are in 
the southern galactic cap $b \leq -40^\circ$ and $\delta \leq
-2.5^\circ$  (1.13 steradians), and 1937 are in the northern
galactic cap  $b \geq 35^\circ$ and $\delta \leq 0^\circ$ (0.57
steradians).  The sample  contains objects (836 galaxies, or 
15 \%  of the sample) previously  observed for the diameter-limited
SSRS  (da Costa \etal 1991).  We measured 2828 new redshifts or 52\% of
the  sample, including galaxies from a survey of the equatorial region
(Huchra \etal 1993).

Tables 2 and 3 contain the SSRS2 catalog.  Table 2 contains 5369
galaxies (3439 in the SSRS2 south and 1930 in the  SSRS2
north) with $m_{SSRS2} \leq 15.5$. These galaxies constitute  a
well-defined, magnitude-limited sample for statistical  analyses. For 
each object the entries are: column (1) the galaxy identification in the
 GSC; column (2) the ESO or MCG identification; columns (3) and (4)  
B1950.0 equatorial coordinates; column (5) the $m_{SSRS2}$ magnitude ; 
column (6) the heliocentric radial velocity $v_\odot$; column (7) the 
estimated internal error, $\epsilon$, in the radial velocity;  column
(8) galaxy  morphologies with T types as discussed in Section 2.3;
column (9) the  source for the radial velocity, where zero represents a
value from the  literature, and  numbers 1 to 9 are the sites in Table
1; column (10) an indicator for galaxies without a radial  velocity
because of  superposed stars (``star''), or because of low surface
brightness requiring long integration times (``lsb''); column (11)
other identifications for galaxies (e.g. NGC or IC number).  Only the 
first page of Table 2 is shown in this publication.  The complete table 
is available in our www site http://obsn.on.br/ssrs2.

Positions are from the GSC, except for those galaxies missed or 
``merged''  into a single object in this catalog.  In these cases,  the
positions, identifications and magnitudes come either from the ESO-LV 
or MCG catalogs,  or are new estimates (16 galaxies mentioned in section
2.1).  For other 6 galaxies we also estimated the magnitudes since the 
quoted value was contaminated by a star image.  Magnitudes estimated for 
these 22 galaxies are labeled with an asterisk.  Sources for radial 
velocities are the same mentioned in da Costa \etal (1991) and for more
recent data, the NASA/IPAC Extragalactic Database (NED).

The completeness in radial velocity for galaxies  with $m_{SSRS2} \le
15.5$ in both galactic caps exceeds 99\%.  We caution users of  this
catalog that there are 35 galaxies without radial velocities.  Of
these, 9 are low surface brightness galaxies and 25 have 
superposed stars which make optical velocities impossible to obtain.

Figure  4 shows the distribution of radial velocities for the SSRS2
 north and south subsamples. To compare the samples in a meaningful way,
we normalize each distribution  by the appropriate survey area.  Both 
subsamples have large amplitude fluctuations.  The subsamples differ
most  noticeably at $\sim$3000 \kms and in the redshift range 8000 -
11000 \kms, where the difference in counts is $\sim 2 \sigma$.  The
clustering  of galaxies explains these differences.  In the south
galactic cap the  great nearby void at 3000 \kms and the Southern wall
at 6000 \kms are  prominent structures; the large overdensity in the
north cap is due to a structure that extends to 8000 \kms. Cone diagrams
for the two surveys are in da Costa \etal (1994a) and da Costa \etal
(1998).

In the early stages of this project (\eg da Costa et al. 1989) we did 
not have a proper calibration of the galaxy magnitudes. We thus took a 
conservative approach when selecting galaxies; we included  galaxies 
from the ESO-LV and MCG catalogs with $m_B \leq 15.5$ as estimated from 
the MCG magnitudes or from $B_T$ in the case of ESO-LV galaxies. As the 
$m_{SSRS2}$ magnitudes became available, some of these objects turned out
to have $m_{SSRS2}$ $>$ 15.5 based on equation (1).  We list these 
57 objects in Table 3, where the columns have the same meaning as 
in Table 2.

\section{Summary}

The SSRS2 catalog, described in this paper, includes 5426 galaxies in 
the southern hemisphere.  For essentially all of these galaxies we
present  radial velocities, good astrometric positions and objectively
measured  magnitudes. The catalog also includes galaxy morphologies
compiled from  a variety of sources.  The magnitude limited sample
designed for  statistical analyses contains 5369 galaxies 
brighter than $m_{SSRS2}$ = 15.5, distributed within a solid angle of
1.7 steradians,  split between two regions in the north and south
galactic caps. The  magnitude system is homogeneous and equivalent to 
isophotal, measured at  26 mag arcsec$^{-2}$. Its zero-point offset
relative to the Zwicky  magnitudes, used in the CfA surveys, is estimated
to be about 0.1$^m$.

Since the loss of objects at faint magnitudes, caused by 
misclassifications of galaxies as stars in the source catalog (GSC),
is small, we conclude that the magnitude limited catalog is essentially
complete to $m_{SSRS2}$ = 15.5.  The redshift completeness exceeds
99\%.  The SSRS2 galaxy catalog is the basis for a variety of studies of
large-scale structure either by itself or in combination with the CfA2
to  probe larger scales.  Together these surveys are the largest
database of  redshifts of the Local Universe.

The authors thank the time allocation committees of all observatories,
for granting this project a generous amount of dark time over a number
of years. More recent observations were also made possible thanks to the
ESO/Observat\'orio Nacional agreement for shared operation of the 1,52 m
ESO telescope at La Silla. The authors thank S. Paoloantonio, C. Valotto
and J.H. Calder\'on for their help in the observations at CASLEO, and
M.A. Nunes and D. Nascimento for their help maintaining the ON Reticon
in working order. We also thank Will Sutherland, Steve Maddox and Somak
Raychaudhuri for providing data from APM surveys.  At the CfA, we thank
Barbara Elwell for her help in the visual inspection stage of the
catalog construction, Susan Tokarz for spectroscopic data reduction, and
Cathy Clemens for her help in the use of the ZCAT database.

LNdC appreciates the hospitality of CfA and IAP where part of this
work was carried out, and acknowledges support from a CNPq fellowship,
the Smithsonian Scholarly Program and the John Simon Guggenheim
Foundation. This project has received support from CNPq-NSF
(Brazil-USA) and CNPq-CONICET (Brazil-Argentina) bilateral
agreements. Partial support has also been provided through CNPq grants
301364/86-9, 453488/96-0 (CNAW), 301373/86-8 (PSP). Additional funds
were provided by the Centro Latino-Americano de F\'\i sica (PSP),
FAPERJ (OLC), NSF AST-9023178 (CNAW) and the ESO Visitor Programme
(CNAW).  The authors acknowledge use of the CCD and data acquisition
system supported under U.S. National Science Foundation grant
AST-90-15827 to R.M. Rich.  This work has made extensive use of the
NASA/IPAC Extragalactic Database (NED) which is operated by the Jet
Propulsion Laboratory, California Institute of Technology, under
contract with the National Aeronautics and Space Administration and of
the NASA Astrophysics Data System.  This work has also used images from
the Digitized Sky Survey,  produced at the Space Telescope Science 
Institute under U.S. Government grant NAG W-2166. The images of these 
surveys are based on photographic data obtained using the Oschin Schmidt 
Telescope on Palomar Mountain and the UK Schmidt Telescope.

\clearpage

%
%
\begin{figure}
\vspace{210mm}
\includegraphics{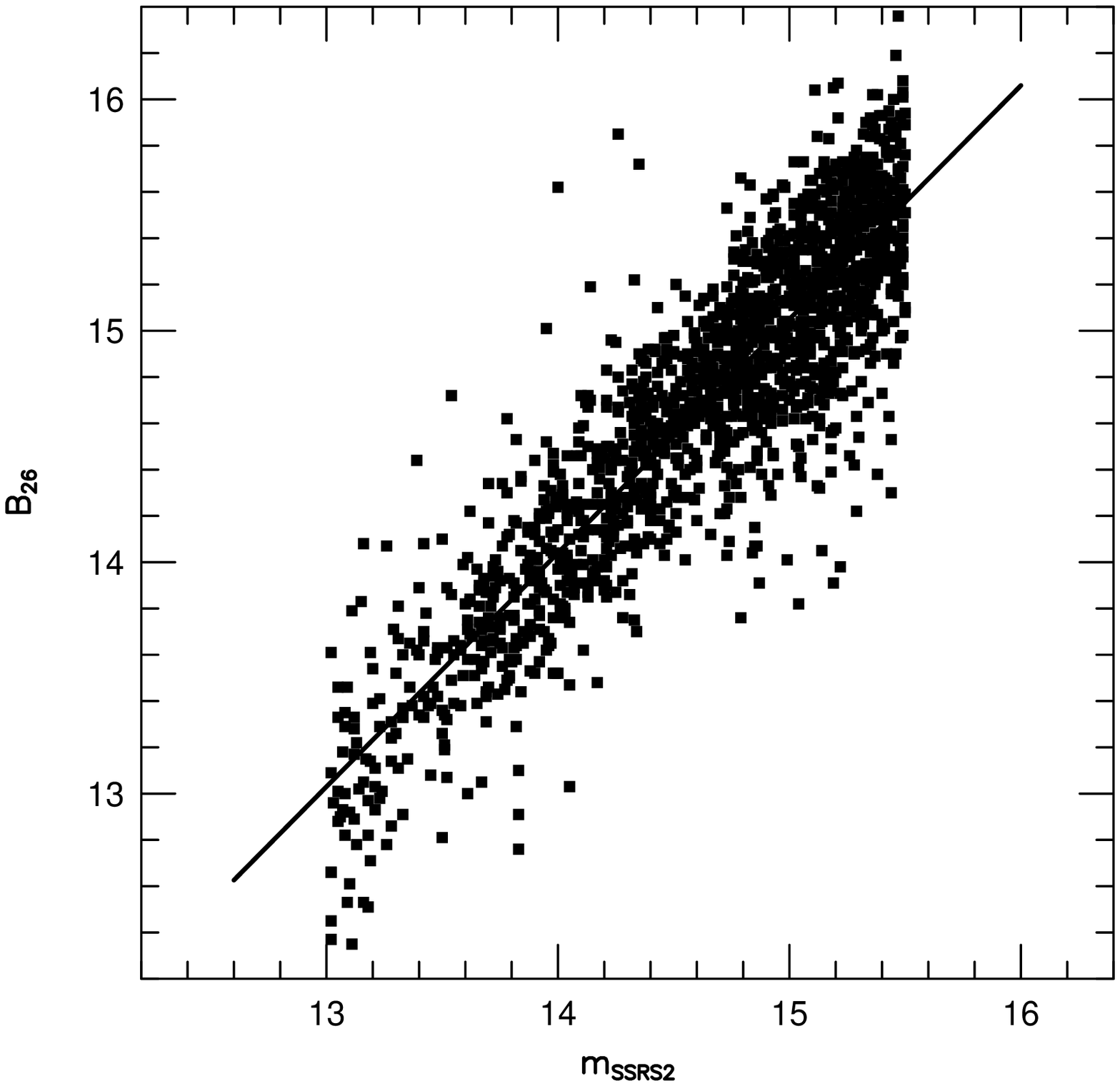}
\caption{Comparison of galaxy magnitudes in our catalog, $m_{SSRS2}$, 
with $B_{26}$ from the ESO-LV catalog for 1588 objects in common between 
the two data sets. The line represents a linear square fit to the data 
points as discussed in the text. }
\end{figure}
\clearpage
%
%
\begin{figure} 
\vspace{210mm}
\includegraphics{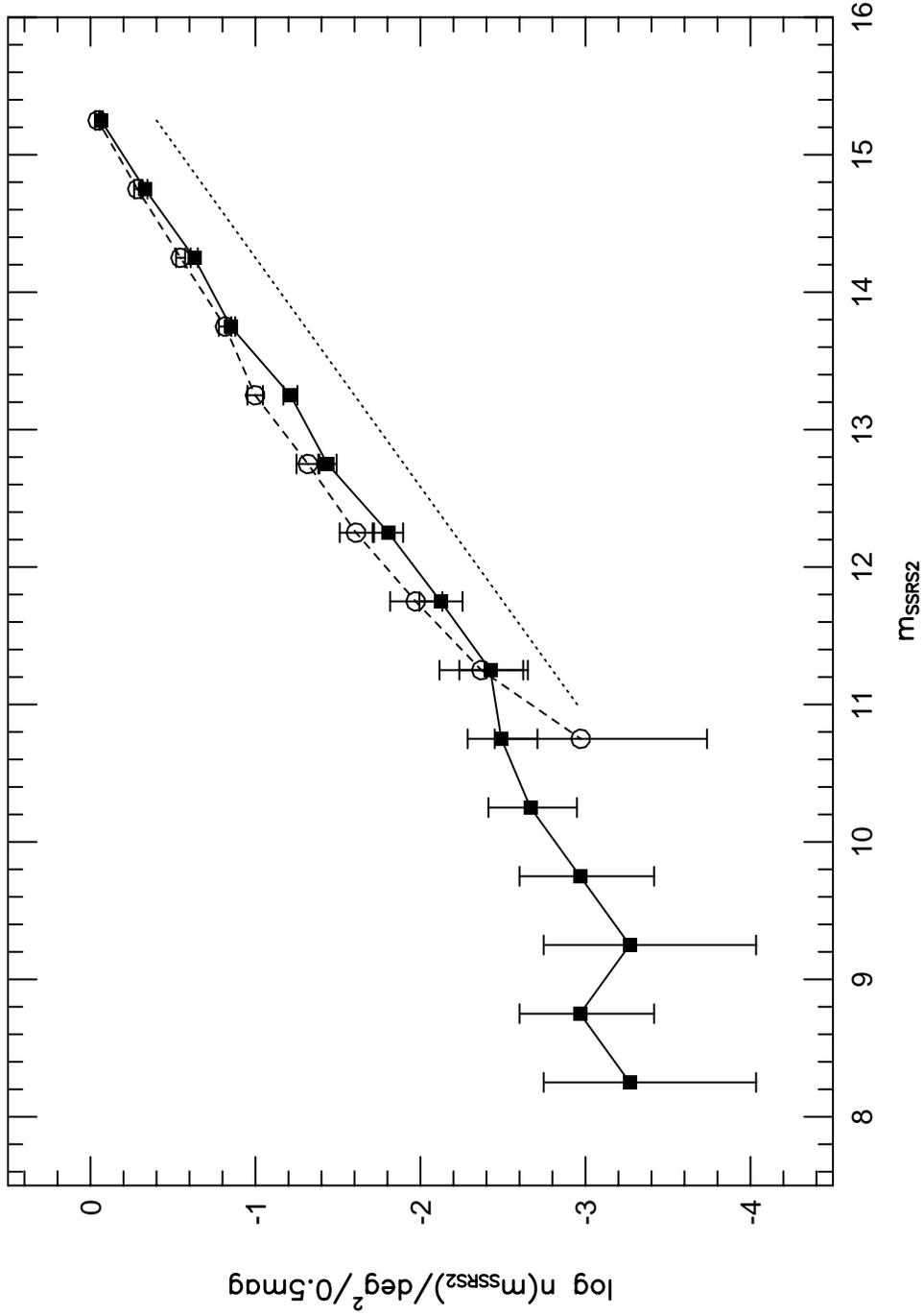}
 \caption{Number counts in galaxies per square degree in
0.5$^m$ bins for the SSRS2 south (squares) and north (circles)
subsamples.  For comparison we show a line with slope equal to 0.6 expected
for a homogeneous distribution with an arbitrary normalization. } 
\end{figure}
\clearpage
%
%
\begin{figure} 
\vspace{210mm} 
\includegraphics{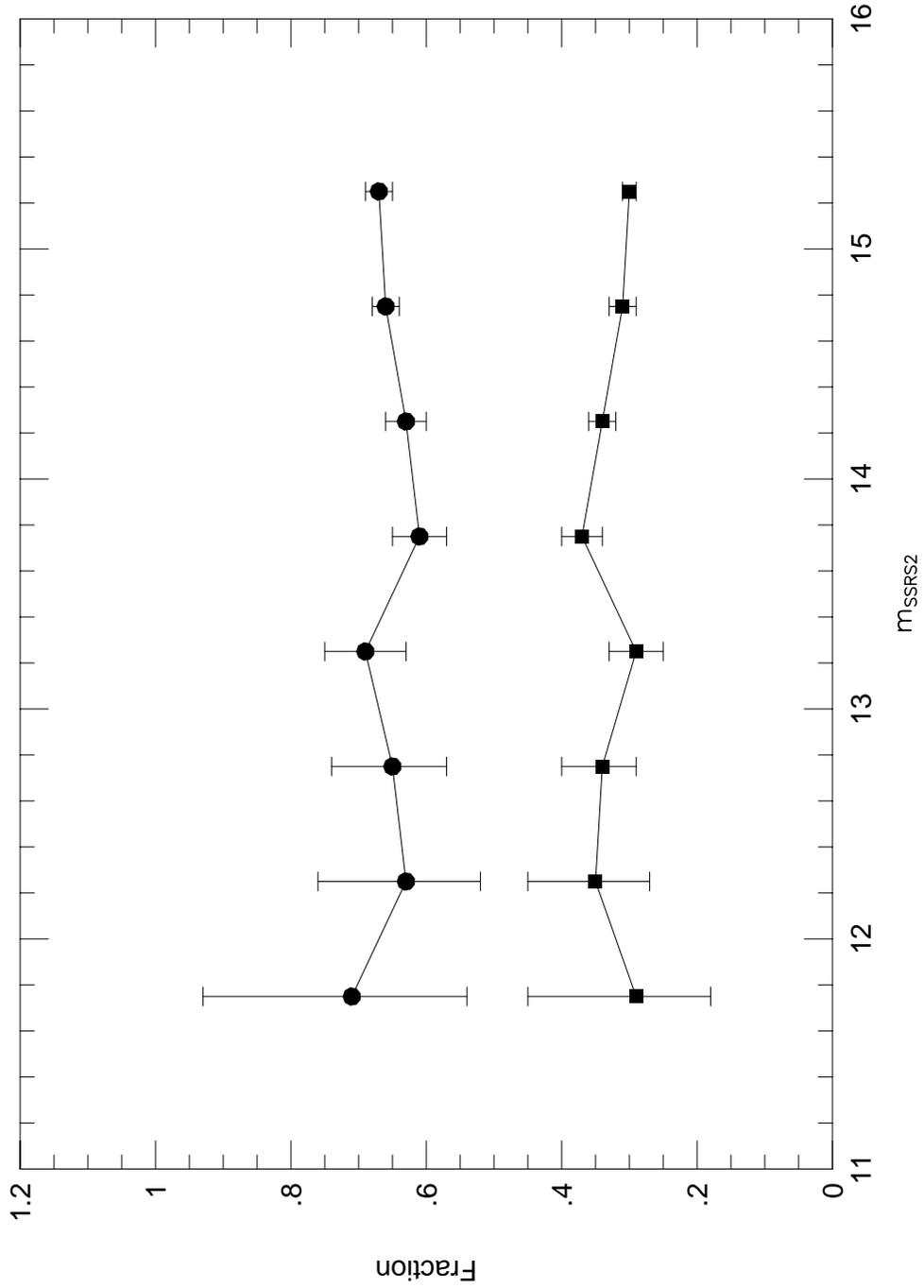} 
\caption{ Distribution in 0.5$^m$ bins of the fraction of different 
morphologies in the SSRS2. Late type galaxies ($1 \le$ T $\le 9$) are 
shown as full circles,  while early--type galaxies ($-5 \le $ T $ \le $0)
are shown as full squares. }
\end{figure} 
\clearpage
%
%
\begin{figure} 
\vspace{210mm}
\includegraphics{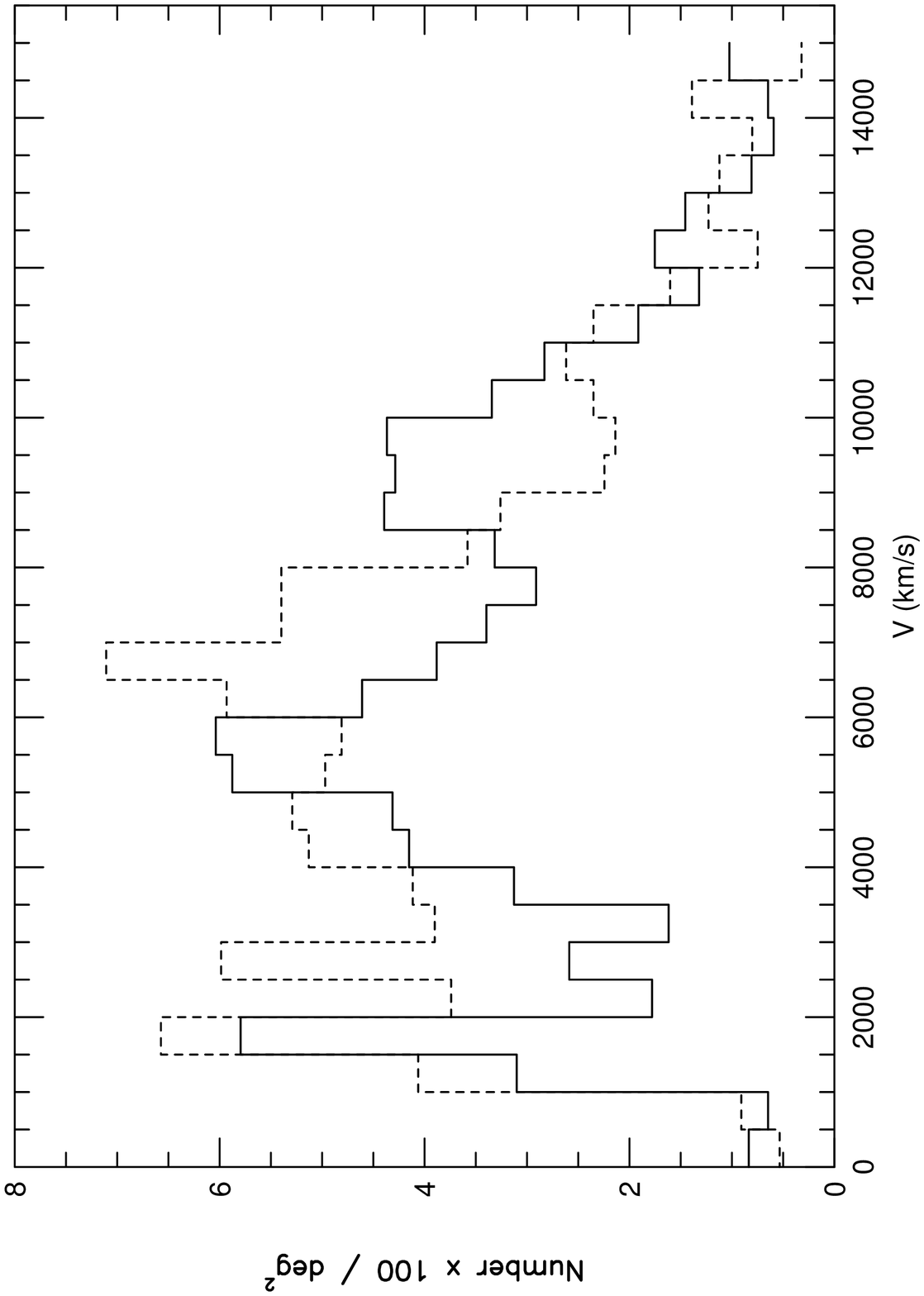} 
\caption{ Distribution of the SSRS2 subsamples in radial
velocity using 500 \kms bins.  The solid line represents the SSRS2 south
sample while the dashed line represents the SSRS2 north. } 
\end{figure}


\clearpage
\begin{references}
\normalsize

\reference {} Alonso, M. V., da Costa, L. N., Pellegrini, P. S., \&
Kurtz, M. J.  1993, AJ, 106, 676

\reference {} Alonso, M. V., da Costa, L. N., Latham, D. W., Pellegrini,
P. S., \& Milone, A. E.  1994, AJ, 108, 1987

\reference {} Barton, E., Geller, M. J., Ramella, M., Marzke, R. O., \&
da Costa, L. N.  1996, AJ , 112, 817

\reference {} Benoist, C., Maurogordato, S., da Costa, L. N., Cappi,
A., \& Schaeffer, R.  1996, ApJ 472, 452

\reference {} da Costa, L. N., Geller, M. J., Pellegrini,
P. S., Latham, D. W., Fairall, A. P., Marzke, R. O., Willmer, C. N. A.,
Huchra, J. P., Calderon, J. H., Ramella, M. \& Kurtz, M. J., 1994a,
 ApJ Letters, 424, L1

\reference {} da Costa, L. N., Pellegrini, P. S., Davis, M., Meiksin,
A., Sargent, W. L. W., \& Tonry, J. L. 1991 ApJS 75, 935

\reference {} da Costa, L. N., Pellegrini, P. S., Nunes, M. A., Willmer,
C. \& Latham, D. W. 1984, AJ 89, 1310

\reference {} da Costa, L. N., Pellegrini, P. S., Sargent, W. L. M.,
Tonry, J., Davis, M., Meiksin, A., Latham, D. W., Menzies, J. W., \&
Coulson, I. A. 1988, ApJ, 327, 544 

\reference {} da Costa, L. N., Pellegrini, P. S., Willmer, C., \&
Latham, D.W.  1989, \apj, 344, 20

\reference{} da Costa, L. N., Vogeley, M., Geller, M. J., Huchra, J. \&
Park, C., 1994b, ApJ, 437, 1

\reference {} Davis, M., Huchra, J., Latham, D. W., \& Tonry, J. 1982,
ApJ, 253, 423

\reference {} da Lapparent, V., Geller, M. J., \& Huchra, J. P. 1986,
ApJ, 302, L1

\reference {}de Vaucouleurs, G., de Vaucouleurs, A.  1964, {\it{ 
Reference Catalogue of Bright Galaxies}}, (Austin: Univ. of Texas Press)

\reference {} El-ad, H., E.-A., Piran, T., \& da Costa, L. N.  1996,
ApJ, 462, L13

\reference {} Fairall, A. P., Willmer, C. N. A., Calder\'on, J. H.,
Latham, D. W., da Costa, L. N., Pellegrini, P. S., Nunes, M. A., Focardi,
P., \& Vettolani, G.  1992, AJ, 103, 11

\reference {} Felten, J. E. 1985, Comments Astrophys., 11, 53

\reference {} Geller, M. J., \& Huchra, J. P. 1989, Science, 246, 897

\reference {} Huchra, J. P.  1976, AJ 81, 952 

\reference {} Huchra, J. P.  1996, private communication (ZCAT)

\reference {} Huchra, J. P., Latham, D. W., da Costa, L. N.,
Pellegrini, P. S., \& Willmer, C. N. A.  1993, AJ, 105, 1637

\reference {} Kurtz, M. J. and Mink, D. J. 1998, submited to PASP,
astro-ph/9803252

\reference {} Kurtz, M. J., Mink, D.J ., Wyatt, W. F., Fabricant, D. G.,
Torres, G., Kriss, G. A., \& Tonry, J. L. 1992, in ASP Conf. Ser. Vol. 25,
Proc. 1$^{st}$ Ann. Conf. Astronomical Data Analysis Software and
Systems, ed., D. M. Worral, C. Biemesderfer, \& J. Barnes
(San Francisco: ASP), 432

\reference {} Lahav, O., et al. 1995, Science, 267, 859

\reference {} Lasker, B. M., Sturch, C. R., McLean, B. M., Russel, J. L.,
Jenker, H., \& Shara, M. 1990, \aj, 99, 2019 (GSC)

\reference{}Lauberts, A. 1982, The ESO/Uppsala Catalogue of the ESO
Quick Blue Survey, (Garching: ESO)

\reference {} Lauberts, A., \& Valentijn, E. A.  1989, The Surface
Photometry Catalogue of the ESO-Uppsala Galaxies, (Garching: ESO) (ESO-LV)

\reference {} Loveday, J., 1996, MNRAS, 278, 1025

\reference {} Marzke, R. O., \& da Costa, L. N.  1997, AJ, 113, 185

\reference {} Marzke, R. O., \& da Costa, L. N., Pellegrini, P. S., \&
Willmer, C. N. A.  1998, Ap.J., {\it in press}

\reference {} Marzke, R. O., Geller, M. J., Huchra, J. P., Corwin,
H. G.  1994$a$, AJ, 108, 437

\reference {} Marzke, R. O., Geller, M. J., da Costa, L. N., \& Huchra,
J. P.  1995, AJ, 110, 477

\reference {} Marzke, R. O., Huchra, J. P., \&  Geller, M. J.,
1994$b$, ApJ 428, 43

\reference {} Massey, P. 1992, A User's Guide to CCD Reductions with
IRAF (Tucson: KPNO Computer Support Group)

\reference {} Park, C., Vogeley, M. S., Geller, M. J., \& Huchra,
J. P.  1994, ApJ, 431, 569

\reference {} Pellegrini, P. S., da Costa, L. N., Huchra, J .P., Latham,
D. W., \& Willmer, C.  1990a., AJ 99, 751

\reference {} Pellegrini, P. S., Willmer, C. N. A., da Costa, L. N., \&
Santiago, B .X.  1990b, ApJ, 350, 95

\reference {} Raychaudhury, S., Lynden-Bell, D., Scharf, C. \& Hudson,
M. J. 1994, Bull American Astron. Soc., 184, \# 38.06

\reference {} Ramella, M., Geller, M. J., \& Huchra, J. P.  1992,
\apj, 384, 396

\reference {} Ramella, M., Pisani, A., \& Geller, M. J. 1997, AJ,
113, 483

\reference {} Ramella, M., Girardi, M., da Costa, L. N., \& Geller, M. J.
1998, in preparation

\reference {} Schectman, S. A., 1996, Landy, S. D., Oemler, A.,
Tucker, D. L., Lin, H., Kirshner, R. P., \& Schechter, P. L.  1996,
ApJS, 47, 172

\reference {} Tonry, J., \& Davis, M. 1979, \aj, 84, 1511

\reference {} Vorontsov-Velyaminov, B. A., \& Arkhipova, V. P.
1963-1968, {\it{The Morphological Catalogue of Galaxies}}, (Moscow:
Moscow University Press), Parts 2-4 (MCG)

\reference {} Willmer, C. N. A., da Costa, L. N., Pellegrini, P. S.,
Latham, D. W., \& Freudling, W.  1995, AJ, 109, 61

\reference {} Willmer, C. N. A., da Costa, L. N., Pellegrini, P. S.  1998,
AJ, 115, 869 

\reference {} Zwicky, F., Herzog, E., Wild, P., Karpowicz, M, \&
Kowal, C. T.  1961-1968,{\it{Catalogue of Galaxies and Clusters of
Galaxies}}, (Pasadena: Caltech), vol I-VI

\end{references}
\end{document}